\documentclass[12pt]{JHEP3}

\usepackage{amsmath}
\usepackage{amsfonts}

\newcommand{\e}{\epsilon}

\renewcommand{\L}{{\mathcal{L}}}
\newcommand{\bL}{\bar{{\mathcal{L}}}}
\newcommand{\z}{{\bar z}}

\renewcommand\O{{\mathcal{O}}}
\newcommand{\be}[1]{ \begin{equation}\label{#1} }
\newcommand{\ee}{\end{equation}}
\newcommand{\ben}[1]{\begin{eqnarray}\label{#1} }
\newcommand{\een}{\end{eqnarray}}
\newcommand{\eq}[1]{(\ref{#1})}
\newcommand{\p}{\partial}

\newcommand{\refb}[1]{(\ref{#1})}
\usepackage{amssymb,amsmath}
\usepackage{epsfig}
\usepackage{epstopdf}
\usepackage{latexsym}
\usepackage{graphicx}
\usepackage{subfigure}
\usepackage{yfonts}

\newcommand{\Hn}{{\mathcal{H}}}
\newcommand{\Gn}{{\mathcal{G}}}
\newcommand{\Mn}{{\mathcal{M}}}

\newcommand{\hn}{{\textswab{h}}}
\newcommand{\gn}{{\textswab{g}}}
\newcommand{\mn}{{\textswab{m}}}

\title{Metrics with Galilean Conformal Isometry}

\author{Arjun Bagchi $^{1}$, Arnab Kundu $^{2}$ \\

$^1$ $\,$School of Mathematics, \\
$\;$ $\,$University of Edinburgh \\
$\;$ $\,$Edinburgh EH9 3JZ, UK. \\

$^2$ $\,$ Theory Group \\
$\;$ $\,$Department of Physics \\
$\;$ $\,$University of Texas at Austin \\
$\;$ $\,$Austin, TX 78712, USA.

$\;$\email{arjun.bagchi@ed.ac.uk, arnab@physics.utexas.edu} 
}

\preprint{EMPG-10-22, UTTG-16-10}

\abstract{The Galilean Conformal Algebra (GCA) arises in taking the non-relativistic limit of the symmetries of a relativistic 
Conformal Field Theory in any dimensions. It is known to be infinite-dimensional in all spacetime dimensions. In particular, the 
2d GCA emerges out of a scaling limit of linear combinations of two copies of the Virasoro algebra. In this paper, we find 
metrics in dimensions greater than two which realize the finite 2d GCA (the global part of the infinite algebra) 
as their isometry by systematically looking at a 
construction in terms of cosets of this finite algebra. We list all possible sub-algebras consistent with some 
physical considerations motivated by earlier work in this direction and construct all possible higher dimensional 
non-degenerate metrics. We briefly study the properties of the metrics obtained. 
In the standard one higher dimensional ``holographic'' setting, 
we find that the only non-degenerate metric is Minkowskian. In four and five dimensions, we find families of 
non-trivial metrics with a rather exotic signature. A curious feature of these metrics is that all but one of them are Ricci-scalar flat. }

\begin{document}

\tableofcontents

\section{Introduction}

Non-relativistic conformal theories have received a lot of recent attention in connection with the AdS/CFT conjecture, more generally the gauge-gravity duality. 
The most popular of the versions of this non-relativistic gauge-gravity duality has been the one studied in the context 
of the Schrodinger algebra. The Schrodinger algebra is the largest symmetry algebra of the free Schrodinger equation 
\cite{Hagen:1972pd, Niederer:1972zz} and has been observed in cold atom systems at unitarity \cite{Nishida:2007pj}. 
Gravity duals of a certain class of field theories possessing Schrodinger symmetry
have been proposed in \cite{Son:2008ye, Balasubramanian:2008dm} and now there is an extensive literature in this line of research, 
some of which can be found in the excellent review \cite{Hartnoll}. Another popular venue of research in this field has been in relation to 
spacetime with Lifshitz symmetry proposed in \cite{Kachru:2008yh}, which unlike the Schrodinger case, does not exhibit invariance under Galilean boosts and hence does not contain the Galilean group as a part of the symmetry algebra.

In \cite{Bagchi:2009my}, a different direction to non-relativistic AdS/CFT was proposed by focusing on a systematic limiting 
procedure of the relativistic symmetry group. The relativistic conformal algebra on the boundary was parametrically contracted to what was called the Galilean 
Conformal Algebra (GCA). One of the remarkable observations here was that the GCA could be given an infinite dimensional lift for 
any spacetime dimensions. It was also observed that the GCA was important to the study of non-relativistic hydrodynamics. 
Specifically, the finite dimensional GCA is the symmetry algebra of the Euler equations, which is valid in cases where the 
fluid viscosity can be neglected. There have been further studies of the various aspects of the GCA in \cite{Alishahiha:2009np} -- \cite{OtherGCA}.

The gravity dual of the GCA was proposed initially to be a novel Newton-Cartan like $AdS_2 \times R^d$ in \cite{Bagchi:2009my}. 
The systematic limit when performed on the parent $AdS$ metric leads to a degeneration and hence the proposal was that when one looked for a 
standard one dimension higher holographic construction, there would be no non-degenerate space-time metric 
and the theory described in terms of connections would be a geometrized version of Newtonian gravity. We should, at this point, 
remind the reader that in the case of the Schrodinger algebra, the gravity dual was found in a two-dimensional higher space-time. 
The question of finding a metric with the Galilean Conformal isometry in higher dimensions remained. Recently, in \cite{Bagchi:2010eg}, a connection between asymptotically
flat spaces and the GCA has been established. The 2d infinite dimensional GCA was shown to be isomorphic to the Bondi-Metzner-Sachs (BMS) algebra \cite{BMS}
in 3 dimensions which is the group of asymptotic isometries of flat three dimensional space 
at null infinity \cite{Barnich:2006av}. The two different points of view are seemingly at loggerheads and one of the issues that we address in this paper is this apparent confusion.

The basic philosophy behind constructing the gravity duals of non-relativistic field theories is to realize the corresponding symmetry group as the isometry group of a spacetime metric. We attempt to find all possible higher (greater than two) dimensional metrics possessing the Galilean Conformal isometry by a process of coset 
construction. In the context of non-relativistic Gauge-gravity duality, the authors of \cite{SchaferNameki:2009xr} have shown that under some ``physical'' 
conditions the metrics obtained by this method uniquely reproduces the holographic constructions with Schr\"{o}dinger and Lifshitz isometries. This procedure has also 
been followed in \cite{Jottar:2010vp}, in relation to the aging algebra, an algebra of relevance to some non-equilibrium statistical mechanical 
systems without time translation symmetry. We conduct a case by case exhaustive study of all possible metrics that can arise out of this coset 
construction for the 2d GCA, using the finite part of the algebra. We look to implement the two ``physical'' conditions as outlined in 
\cite{SchaferNameki:2009xr} and then make our search more extensive by relaxing one of them. We find that when we are looking at metrics 
with one extra direction, the ``physical'' conditions do not lead to any non-degenerate spacetime metric in 3-d, adding strength 
to the claim that the correct structure to look for is indeed a Newton-Cartan like $AdS_2 \times R$. Interestingly, when one of 
the two ``physical'' conditions are relaxed, we obtain a flat 3d metric in keeping with the connection discussed in \cite{Bagchi:2010eg}. We find other 
non-degenerate metrics for higher dimensional spaces. Curiously, most of these metrics turn out to be Ricci-scalar flat, although
(except for the Minkowskian one) they source non-trivial Ricci tensors. 

The outline of the paper is as follows: we first review, in Sec.~2, the Galilean Conformal Algebra with special emphasis on 
the 2d GCA which shall be the focus of the paper. In Sec.~3, we outline the procedure of constructing metrics on 
homogeneous coset spaces that we would use. Sec.~4 contains the main results of the paper. We sub-divide the section
according to the dimension of the space-time metric that we construct and make several comments. The main results are 
also summarized in a table in this section. We end with some concluding remarks. An appendix contains a list of all
possible sub-algebras for the 2d GCA.

\section{A Review of the GCA}

\subsection{GCA in arbitrary dimensions}

The maximal set of conformal isometries of Galilean spacetime generates the infinite dimensional Galilean Conformal
Algebra \cite{Bagchi:2009my}. The notion of Galilean spacetime is a little subtle since the spacetime metric degenerates 
into a spatial part and a temporal piece. Nevertheless there is a definite limiting sense (of the relativistic spacetime) 
in which one can define the conformal isometries (see \cite{Duval:2009vt}) of the nonrelativistic geometry. Algebraically, 
the set of vector fields generating these symmetries 
are given by
\ben{gcavec}
L^{(n)} &=& -(n+1)t^nx_i\p_i -t^{n+1}\p_t \,,\cr
M_i^{(n)} &=& t^{n+1}\p_i\,, \cr
J_a^{(n)} \equiv J_{ij}^{(n)} &= & -t^n(x_i\p_j-x_j\p_i)\,,
\een 
for integer values of $n$. Here $i=1\ldots (d-1)$ range over the spatial directions. 
These vector fields obey the algebra
\ben{vkmalg}
[L^{(m)}, L^{(n)}] &=& (m-n)L^{(m+n)}, \qquad [L^{(m)}, J_{a}^{(n)}] = -n J_{a}^{(m+n)}, \cr
[J_a^{(n)}, J_b^{(m)}]&=& f_{abc}J_c^{(n+m)}, \qquad  [L^{(m)}, M_i^{(n)}] =(m-n)M_i^{(m+n)}. 
\een
There is a  finite dimensional subalgebra  of the GCA (also sometimes
referred to as the GCA) which consists of taking $n=0,\pm1$ for the
$L^{(n)}, M_i^{(n)}$ together with $J_a^{(0)}$. This algebra is obtained
by considering the nonrelativistic  contraction of the usual (finite
dimensional) global conformal algebra $SO(d,2)$ (in $d>2$ spacetime
dimensions) (see for example \cite{Bagchi:2009my, Lukierski:2005xy}).

\subsection{GCA in 2d}

In two spacetime dimensions, as is well known, the situation is special. The relativistic conformal algebra is infinite dimensional
and consists of two copies of the Virasoro algebra. 
One expects this to be related to the infinite dimensional GCA algebra \cite{Bagchi:2009}. 
In two dimensions the non-trivial generators in  \eq{vkmalg} are the $L_n$ and the $M_n$:
\be{gca2dvec}
L^{(n)} = -(n+1)t^n x\p_x -t^{n+1}\p_t\,, \quad M^{(n)} = t^{n+1}\p_x\,,
\ee
which obey
\be{vkmalg2d}
[L^{(m)}, L^{(n)}] = (m-n)L^{(m+n)}\,, \quad [M^{(m)}, M^{(n)}] =0\,, \quad [L^{(m)}, M^{(n)}] = (m-n) M^{(m+n)} \ .
\ee

These generators  in \eq{gca2dvec} arise precisely from a nonrelativistic contraction of the two copies of the Virasoro algebra.
To see this, let us remember that the non-relativistic contraction consists of taking the scaling 
\be{nrelscal}
t \rightarrow t\,, \qquad   x \rightarrow \epsilon x\,,
\ee
with $\epsilon \rightarrow 0$. This is equivalent to taking the velocities $v \sim \epsilon$ to zero
(in units where $c=1$). Consider the vector fields which generate (two copies of) the centre-less 
Virasoro Algebra in two dimensions:
\be{repn2dV}
\L^{(n)} = -z^{n+1} \p_z\,, \quad \bL^{(n)} = -\z^{n+1} \p_{\z}\,.
\ee 
In terms of space and time coordinates, $z= t+x$, $\z=t-x$. Expressing $\L_n , \bL_n$ in terms of $t,x$ 
and taking the above scaling \eq{nrelscal} reveals that in the limit the combinations
\be{GCArepn}
\L^{(n)} + \bL^{(n)} = -t^{n+1}\p_t - (n+1)t^nx \p_x + \O(\e^2); \quad \L^{(n)} - \bL^{(n)} = -{1\over \e}t^{n+1} \p_x + \O(\e)\,.
\ee 
Therefore we see that as $\e\rightarrow 0$ \cite{Bagchi:2009}
\be{Vir2GCA}
\L^{(n)} + \bL^{(n)} \longrightarrow L^{(n)}\,, \quad \e (\L^{(n)} - \bL^{(n)}) \longrightarrow - M^{(n)}\,.
\ee

Let us now rewrite the $(1+1)$-dimensional (finite) algebra generated by $\{L^{(\pm 1)}, L^{(0)}\}$ and $\{M^{(\pm 1)}, M^{(0)}\}$.
The non-trivial commutators resulting from (\ref{vkmalg2d}) are given by
\begin{eqnarray}\label{gca2d}
&& \left[ D, H \right] = H \ , \quad \left[ D, K_0 \right] = - K_0 \ , \quad \left[ D, K_1 \right] = - K_1 \ , \quad \left[ D, P \right] = P \ , \nonumber\\
&& \left[ K_0, H \right] = 2 D \ , \quad \left[ B, H \right] = P \ , \quad \left[ K_1, H \right] = 2 B \ , \nonumber\\
&& \left[ K_0, B \right] = K_1 \ , \quad \left[ K_0, P \right] = 2 B \ ,
\end{eqnarray}
where we have made the following identifications
\begin{eqnarray}
&& L^{(-1)} \equiv H \ , \quad L^{(0)} \equiv D \ , \quad L^{(+1)} \equiv K_0 \ , \nonumber\\
&& M^{(-1)} \equiv P \ , \quad M^{(0)} \equiv B \ , \quad M^{(+1)} \equiv K_1 \ .
\end{eqnarray}
Here $H$ is the time translation generator, $D$ is the dilatation operator, $P$ is the spatial translation generator, $B$ is the Galilean boost and $K_0, K_1$ are the two components of the special conformal generator. These identifications naturally arise when one considers the contraction of the relativistic conformal algebra 
\cite{Bagchi:2009my}. In the rest of the paper we will entirely focus on the algebra written in (\ref{gca2d}) and not
be concerned about the infinite dimensional extension. We would look to realize this finite algebra as the isometries
of spacetime metrics in dimensions greater than two. It is natural to expect that only the finite GCA would 
play the role of the true isometries and the other higher modes may correspond to asymptotic isometries of the 
metrics that we would obtain\footnote{For a brief review on asymptotic isometries see {\it e.g.} \cite{Compere:2009qm}.}. This is something that we would not address in the current paper.

\section{Construction of Metrics on Coset Spaces}

Here we briefly review the construction of metrics on coset spaces that we will use in the rest of the paper. 
We closely follow the notation and conventions of \cite{SchaferNameki:2009xr}. We would like to consider a coset $\Mn = \Gn / \Hn$, where $\Gn$ is the Galilean 
Conformal group and $\Hn$ is a subgroup of $\Gn$. The corresponding Lie algebras are denoted by $\gn$ and $\hn$ 
respectively and for each $g \in \gn$ there is a corresponding element denoted by $[g] \in \gn / \hn$. As vector spaces, we can always decompose
\begin{eqnarray}\label{vecdecom}
\gn = \hn \oplus \mn \ .
\end{eqnarray}
The coset $\Mn$ is called a reductive coset if there exists a choice of $\mn \in \Mn$ such that
 $[\hn, \mn] \subset \mn$. We will see that for the GCA generically we do not have such reductive cosets.

Our goal here will be to construct a $\Gn$-invariant metric on the homogeneous space $\Mn$. 
Given a Lie group the Cartan-Killing form is given by
\begin{eqnarray}
\Omega_{ab} \equiv \frac{1}{I_{\rm adj}} f_{ac}^{\ \  d} f_{bd}^{\ \ c} \ ,
\end{eqnarray}
where $f_{ab}^{\ \ c}$ are the structure constants and $I_{\rm adj}$ is the Dynkin index. 
For a semi-simple Lie group the Cartan-Killing form in non-degenerate and therefore induces 
a non-degenerate $\Gn$-invariant metric on $\Mn$. However, the GCA is not a semi-simple algebra 
and thus the corresponding Killing form is degenerate. We would like to point out here that there is the possibility of 
constructing a non-degenerate two-form over the whole group manifold {\it via} a procedure called ``double extension". We would have more to say about this later.

Following \cite{KN}, there exists a one-to-one correspondence between $\Gn$-invariant metric on 
$\Mn = \Gn / \Hn$ and Ad($\Hn$)-invariant non-degenerate 
symmetric bilinear forms $\Omega$ on $\gn/ \hn$. When $\Hn$ is connected, this invariance takes the following form
\begin{eqnarray}\label{omega}
\Omega_{[m][n]} f_{[k] p} ^{\ \ \ [m]} + \Omega_{[k][m]} f_{[n] p }^{\ \ \ [m]} = 0 \ ,
\end{eqnarray}
where $[m], [n], \ldots$ are indices corresponding to $\mn$ and $p$ indicates the index corresponding to $\hn$. 
Given the structure constants for a particular choice of $\hn$ and $\mn$, we can solve for the bilinear $\Omega$ 
from this equation.

However, the existence or the uniqueness of a solution for $\Omega$ is not guaranteed and we will observe later 
that for the GCA only a few choices for the sub-algebra $\hn$ we have a non-degenerate $\Omega$. Moreover, a typical 
solution of (\ref{omega}) does not fix $\Omega$ completely, rather gives a symmetric bilinear in terms of a bunch of 
arbitrary real numbers. This therefore will result in redundancies in the description of the $\Gn$-invariant 
(family of) metrics that we will eventually obtain.

Now let us choose an explicit coordinate basis as in \cite{SchaferNameki:2009xr}. First we fix a linear space 
decomposition (\ref{vecdecom}) and denote that $t_m, t_n, \ldots$ are the basis of $\hn$ and $t_p, t_q, \ldots$ 
are the basis of $\mn$. 
Then an element $[g] \in \Gn/ \Hn$ can be represented by
\begin{eqnarray}
[g] = \left[{\rm exp} \left(x_m t_m \right) {\rm exp} \left(x_n t_n \right) \ldots \right] \quad {\rm modulo} \, 
\Hn \ .
\end{eqnarray}
The Maurer-Cartan one-form given by $J_g = g^{-1} d g$ can then be computed according to the linear space 
decomposition in (\ref{vecdecom})
\begin{eqnarray}
J_g = e_m t_m + e_p t_p \ ,
\end{eqnarray}
where $e_m$ and $e_p$ are the vielbein. The metric on the coset is then constructed by contracting the symmetric 
bilinear $\Omega$ with the vielbein
\begin{eqnarray}\label{metric}
G = \Omega^{pq} e_p e_q \ . 
\end{eqnarray}
%

\section{Homogeneous spaces with $2$d Galilean Conformal isometry}

In this section we will discuss and present the non-trivial homogeneous spaces 
(and the corresponding choice of the sub-algebra) that we obtain {\it via} the coset construction. 
For the interested reader, we have presented a  complete list of all possible sub-algebras of the $2$d GCA in appendix \ref{sub}.

Let us mention our guiding principles for the choices of sub-algebra here. In \cite{SchaferNameki:2009xr}, 
the authors uniquely determined the metrics for the Schrodinger and the Lifshitz algebras by imposing the following 
``physical" conditions:
\begin{enumerate}
 \item $\hn$ does not contain the translation generator $P$.
 \item $\hn$ contains the boost generator $B$. 
\end{enumerate}

As argued in \cite{SchaferNameki:2009xr}, condition (1) is natural in the sense that $P$ would induce infinitesimal translations in the resulting geometry and
should not be included in the stabilizer of a point in $\Gn/\Hn$. We shall strictly follow condition (1) in all our
examples. Condition (2) is derived from the higher dimensional analogue of Lorentz invariance. For a $d$ dimensional
algebra, the authors of \cite{SchaferNameki:2009xr} proposed to keep $J_{ij}, B_i$ in $\hn$ to preserve Lorentz invariance
in $d$ dimensions. We, however, do not believe in the sanctity of this condition in our analysis and would proceed to 
relax it in our exhaustive study.

\subsection{$3$-dimensional Minkowski space}

We begin by considering the case when dim$\Mn = 3$. In this case, the only choice that gives a non-degenerate 
symmetric bilinear $\Omega$ (and therefore a non-degenerate metric) is the coset $\Mn = \Gn / \{ H, D, K_0 \}$. 
Note that in this case the sub-algebra does not contain the boost generator $B$ and thus it falls under the category 
where we relax one of the ``physical" conditions outlined above (and in \cite{SchaferNameki:2009xr}).

The structure constants are given by
\begin{eqnarray}
f_{[i] H}^{\ \ \ [j]} = \begin{pmatrix}
  0 & 0 & 0   \\
  0 & 0 & 2  \\
  1 & 0 & 0   
\end{pmatrix} \ , \quad f_{[i] D}^{\ \ \ [j]} = \begin{pmatrix}
  -1 & 0 & 0  \\
  0 & 1 & 0 \\
  0 & 0 & 0  
\end{pmatrix} \ , \quad f_{[i] K_0}^{\ \ \ [j]} = \begin{pmatrix}
  0 & 0 & -2  \\
  0 & 0 & 0 \\
  0 & -1 & 0  
\end{pmatrix} \ ,
\end{eqnarray}
which gives
\begin{eqnarray}
\Omega = \begin{pmatrix}
  0 & -2 \omega_{33} &  0   \\
  -2 \omega_{33} & 0 &  0  \\
   0 & 0 & \omega_{33}
\end{pmatrix} \ , \quad \omega_{ij} \in \mathbb{R} \ .
\end{eqnarray}
Now the coset element is parametrized as
\begin{eqnarray}
[g] = [e^{x_P P} e^{x_1 K_1} e^{x_B B}] \ ,
\end{eqnarray}
which gives the following vielbein
\begin{eqnarray}
e_P = dx_P \ , \quad e_{K_1} = dx_1 \ , \quad e_B = dx_B \ .
\end{eqnarray}
So, the resulting metric reads:
\be{}
ds_3^2 = \Omega^{pq} e_p e_q =  \omega_{33} (- 4 dx_P dx_1 + dx_B) \ .
\ee

This is the flat $3$-d Minkowski space\footnote{Clearly we can set $\omega_{33} = 1$ without any loss of generality.}.
As we remarked earlier based on the observation recently made in \cite{Bagchi:2010eg}, this is a consequence of the 
isomorphism between the finite Galilean Conformal group in $(1+1)$-dimensions and the Poincar\'{e} group in 
$(2+1)$-dimensions. The isomorphism actually extends beyond the finite GCA and encompasses the full infinite extension
of the GCA on one side and the infinite-dimensional BMS group in 3 dimensions which is the asymptotic symmetric group
of flat 3d space-times at null infinity \cite{Bagchi:2010eg}.  

We observe that the strict imposition of both the ``physical conditions'' above does not lead to any non-degenerate
spacetime metric. As remarked in the introduction, the original proposal for the dual gravitational description of 
a system with the GCA was given in terms of a Newton-Cartan like AdS \cite{Bagchi:2009my}. In the case of the three dimensional bulk dual,
the structure of the spacetime would be a fibre bundled $AdS_2 \times R$. The space-time metric degenerates and the 
dynamical quantities are the Chritoffel symbols which ``talk'' to the separate metrics of the base $AdS_2$ and the fibres.
The imposition of Lorentz symmetry in two dimensions (condition (2)) in our present construction rules out a non-degenerate
spacetime metric and this is in keeping with the claim that the correct structure to look for is a Newton-Cartan like
$AdS_2 \times R$.

Let us comment on a couple of things here about the flat metric that we have obtained. Firstly, we know that if 
an $n$-dimensional manifold admits ${1\over 2}n(n+1)$ Killing vectors, it must be a manifold of constant curvature. 
We were looking for spacetimes in 3 dimensions admitting the 6 dimensional GCA as an isometry. 
So, we would have ended up with spacetimes of constant curvature, our only choices are: flat, de-Sitter,
or anti de-Sitter in three dimensions. That we get a flat spacetime is thus not a surprise. 

Another point to note is that this seems to be the metric that is picked out by the method of contractions on that gave
rise to the GCA from the relativistic conformal algebra from the point of view of AdS/CFT \cite{Bagchi:2009my},
both on the boundary and in the bulk. To see this, let us remind ourselves that the $AdS_3$ metric is obtained by the
following coset construction (see {\it e.g.} \cite{Kraus:2006wn}):
\be{ads-coset}
AdS_3 = {{ SL(2, R) \times SL(2,R)}\over {SL(2,R)_{\rm diag}}}  \ .
\ee
The above construction of the Minkowskian metric of the GCA is precisely the contraction of \refb{ads-coset}
{\footnote{We would like to thank Rajesh Gopakumar for pointing this out to us.}}.
The finite GCA is obtained by contracting the global $SL(2, R) \times SL(2,R)$ of the Virasoro algebra and 
${SL(2,R)_{\rm diag}}$, the diagonal $SL(2,R)$ subgroup of the relativistic theory, is parent of the $\{ H, D, K_0 \}$ 
subalgebra of the GCA.

\subsection{$4$-dimensional metrics}

Next we consider the case when dim$\Mn = 4$. The first non-trivial case is the coset $\Mn = \Gn / \{B, D\}$, 
which obeys both the ``physical" conditions outlined in \cite{SchaferNameki:2009xr}. 
In this case the structure constants are given by
\begin{eqnarray}
f_{[i] B}^{\ \ \ [j]} = \begin{pmatrix}
  0 & -1 & 0 & 0  \\
  0 & 0 & 0 & 0  \\
  0 & 0 & 0 & +1  \\
  0 & 0 & 0 & 0  
\end{pmatrix} \ , \quad f_{[i] D}^{\ \ \ [j]} = \begin{pmatrix}
  -1 & 0 & 0 & 0  \\
  0 & -1 & 0 & 0  \\
  0 & 0 & +1 & 0  \\
  0 & 0 & 0 & +1  
\end{pmatrix} \ ,
\end{eqnarray}
which yields
\begin{eqnarray} \label{omega41}
\Omega = \begin{pmatrix}
  0 & 0 &  \omega_{13} &  \omega_{14}  \\
  0 & 0 & \omega_{14} & 0  \\
   \omega_{13} & \omega_{14} & 0 & 0  \\
   \omega_{14} &  0 &  0 &  0  
\end{pmatrix} \ , \quad \omega_{ij} \in \mathbb{R} \ .
\end{eqnarray}
This is non-degenerate as long as $\omega_{14} \not = 0$.

Since we get a non-degenerate bilinear, let us compute the vielbein in this case. We parametrize the coset element as
\begin{eqnarray}
[g] = [e^{x_H H} e^{x_P P} e^{x_0 K_0} e^{x_1 K_1} ] \ ,
\end{eqnarray}
which gives the following vielbein
\begin{eqnarray} \label{viel41}
&& e_H = dx_H \ , \quad e_P = dx_P \ , \nonumber\\
&& e_{K_0} = x_0^2 dx_H  + dx_0 \ , \nonumber\\
&& e_{K_1} = 2 x_0 x_1 dx_H + x_0^2 dx_P + dx_1 \ .
\end{eqnarray}
For the sake of visualization, let us write down the full metric.
We define $x_H = t, x_P = x, x_0=y, x_1 =z, w_{31} =a, w_{41}=b$. The metric, then, can be written as:
\be{4d-1}
ds_{4(1)}^2 = (2 a y^2 + 4 byz) dt^2 + 4 by^2 dt dx + 2 a dt dy + 2 b dt dz + 2 b dx dy \ .
\ee
Note that here we have two arbitrary real numbers $a, b$ which parametrize a family 
of metrics. This family of metrics has vanishing Ricci-scalar.

The other non-trivial result comes from taking the coset $\Mn = \Gn / \{B, \alpha_1 D + \alpha_2 K_1\}$. 
The structure constants are given by
\begin{eqnarray}
f_{[i] B}^{\ \ \ [j]} = \begin{pmatrix}
  0 & -1 & 0 & 0  \\
  0 & 0 & 0 & 0  \\
  0 & 0 & 0 & 0  \\
  0 & 0 & - \alpha_1/ \alpha_2 & 0  
\end{pmatrix} \ , \quad f_{[i] \alpha_1 D + \alpha_2 K_1}^{\ \ \ \ \ \ \ \ \ \ \ \ \ [j]} = \alpha_1 \begin{pmatrix}
  -1 & 0 & 0 & 0  \\
  0 & -1 & 0 & 0  \\
  0 & 0 & +1 & 0  \\
  0 & 0 & 0 & +1  
\end{pmatrix} \ ,
\end{eqnarray}
which yields the following
\begin{eqnarray} \label{omega42}
\Omega = \begin{pmatrix}
  0 & 0 &  \omega_{31} &  \omega_{41}  \\
  0 & 0 & 0 & - \frac{\alpha_1}{\alpha_2} \omega_{31}  \\
  \omega_{31} & 0 & 0 & 0  \\
  \omega_{41} &  - \frac{\alpha_1}{\alpha_2} \omega_{31} &  0 &  0  
\end{pmatrix} \ , \quad \omega_{ij} \in \mathbb{R} \ .
\end{eqnarray}
The above $\Omega$ is non-degenerate for $\alpha_1 \not = 0$. The vielbein are obtained to be:
\begin{eqnarray} \label{viel42}
&& e_H = e^{-x_D} dx_H \ , \quad e_P = e^{- x_D} dx_P \ , \nonumber\\
&& e_D = dx_D - 2 x_0 e^{- x_D} dx_H - \frac{\alpha_1}{\alpha_2} dx_P \ , \nonumber\\
&& e_{K_0} = e^{- x_D} x_0^2 dx_H - x_0 dx_D + dx_0 \ .
\end{eqnarray}
It can be checked that without any loss of generality we can set\footnote{This is achieved by computing the Ricci 
tensor and observing that only the ratios $\omega_{41}/\omega_{31}$ and $\alpha_2/\alpha_1$ appear.} 
$\omega_{31} = 1 = \alpha_2$. Hence we get a family of metrics parametrized by two real numbers $\omega_{41}$ 
and $\alpha_1$. Again, for clarity, it is useful to write the metric down explicitly. We make the following redefinitions:
$x_H = t, x_P =x, x_0 = y, e^{x_D} = r, \omega_{41} = a, \alpha_2 = \alpha$.
\be{4d-2}
ds_{4(2)}^2 = {2 \over r^2} \{  (1-ay) dr dt +  (a y^2 - 2y) dt^2 -  \alpha (r+y^2) dt dx +  ar dt dy +  \alpha y dx dr
- \alpha r dx dy \} \ .
\ee
It is trivial to check that this metric also has vanishing Ricci-scalar. This is the only 
non-reductive example that we encounter in the coset construction of the $2$-dimensional Galilean Conformal symmetry.

Let us offer some comments regarding the signature of these $4$-dimensional metrics. 
It can be observed that the two distinct families of metrics we obtained take the following generic form
\begin{eqnarray} \label{gen4d}
ds^2 = 2 \Omega_{13} e_1 e_3 + 2 \Omega_{14} e_1 e_4 + 2 \Omega_{2,(3/4)} e_2 e_{(3/4)} \ ,
\end{eqnarray}
where $\Omega_{ij}$ are the corresponding matrix entries in (\ref{omega41}) or (\ref{omega42}) and $e_i$'s are 
the vielbein given in (\ref{viel41}) or (\ref{viel42}). If we introduce a local orthonormal frame 
$\{E_1, E_2, E_3, E_4\}$, where $E_i$'s are appropriate linear combinations of $e_i$'s, 
the particular form of the metric in (\ref{gen4d}) is strongly suggestive that the signature of the 
metric should be $(2,2)$.\footnote{It is easy to check that one cannot write $ds^2 = - E_1^2 + E_2^2 + E_3^2 + E_4^2$;
 however one can write $ds^2 = - E_1^2 - E_2^2 + E_3^2 + E_4^2$.} 
It is worth noting at this point that in \cite{Martelli:2009uc}, a geometric realization of the 
``exotic'' Galilean Conformal Isometry in $(2+1)$-dimensions (called ``exotic'' because of the existence of 
a central charge in the commutator of the boost generators on the plane which is special to these dimensions)
 was found in terms of an $AdS_7$-metric with $(3,4)$ signature.

\subsection{$5$-dimensional metrics}

Finally we present the $5$-dimensional metrics obtained {\it via} the coset construction. 
The first non-trivial case is the coset $\Mn = \Gn / \{ B \}$. This gives the following symmetric bilinear
\begin{eqnarray} \label{5d1omega}
\Omega = \begin{pmatrix}
  \omega_{11} & 0 &  \omega_{13} &  \omega_{14} &  \omega_{15} \\
  0 & 0 & 0 &  \omega_{15} & 0 \\
   \omega_{13} & 0 &  \omega_{33} &  \omega_{34} & 0 \\
   \omega_{14} &  \omega_{15} &  \omega_{34} &  \omega_{44} & 0 \\
   \omega_{15} & 0 & 0 & 0 & 0 
\end{pmatrix} \ , \quad \omega_{ij} \in \mathbb{R} \ ,
\end{eqnarray}
which is non-degenerate if $\omega_{15} \not = 0$ and $\omega_{33} \not = 0$ and without any loss of generality 
we can set $\omega_{33} = 1 = \omega_{13} = \omega_{14} = \omega_{15} = \omega_{34}$. In this case we get the 
following vielbein
\begin{eqnarray}\label{5d1viel}
&& e_H = e^{- x_D} dx_H \ , \quad e_P = e^{- x_D} dx_P \ , \quad e_D = - 2 x_0 e^{- x_D} dx_H + dx_D \ , \nonumber\\
&& e_{K_0} = x_0^2 e^{- x_D} dx_H - x_0 dx_D + dx_0 \ , \nonumber\\
&& e_{K_1} = 2 x_0 x_1 e^{- x_D} dx_H + x_0^2 e^{- x_D} dx_P - x_1 dx_D + dx_1 \ .
\end{eqnarray}
The resulting two-parameter family of metrics is Ricci-scalar flat. 
Clearly, this construction obeys both the ``physical" conditions.

The only other non-trivial example in $5$-dimensions is the coset $\Mn = \Gn / \{ D \}$, 
which does not obey the ``physical" condition $(2)$. In this case we get
\begin{eqnarray}\label{5d2omega}
\Omega = \begin{pmatrix}
  0& 0 &  \omega_{31} & \omega_{41} & 0 \\
  0 & 0 &  \omega_{32} & \omega_{42} & 0 \\
  \omega_{31} & \omega_{32} & 0 & 0 & 0 \\
  \omega_{41} & \omega_{42} & 0 & 0 & 0 \\
   0 & 0 & 0 & 0 &  \omega_{55} 
\end{pmatrix} \ , \quad \omega_{ij} \in \mathbb{R} \ .
\end{eqnarray}
This is also non-degenerate provided $\omega_{55} \not = 0$ and $\omega_{32} \omega_{41} \not = \omega_{31} 
\omega_{42} $. The vielbein are given by
\begin{eqnarray}\label{5d2viel}
&& e_H = dx_H \ , \quad e_P = dx_P \ , \quad e_{K_0} = x_0^2 dx_H + dx_0 \ , \nonumber\\
&& e_{K_1} = \left(2 x_0 x_1 + x_0^2 x_B \right) dx_H + x_0^2 dx_P + x_B dx_0 + dx_1 \ , \nonumber\\
&& e_B = - 2 x_0 dx_P + dx_B \ .
\end{eqnarray}
This actually gives a four-parameter family of metrics. 
This family generically has coordinate dependent Ricci scalar which diverges as $R \sim x_1^2$ for $x_1 \to \infty$. 
If $\omega_{32} = 0$, we still get a non-degenerate metric but the Ricci-scalar vanishes identically. 
On the other hand, if $\omega_{32} \not = 0$, then the Ricci-scalar can vanish at a particular point in $x_1$.

As in the examples with $4$-dimensional metrics, it can be also argued that the existence of a local orthonormal
 frame and the precise structure of these $5$-dimensional metrics strongly suggests the signature be $(2,3)$.

Note that the $2$-dimensional GCA has $6$ generators; hence a homogeneous space of $5$-dimensions is constructed
by choosing a sub-algebra which consists of only one generator. This is a rather trivial choice which nonetheless
yields a family of non-trivial metrics.

Finally, we summarize some of our results in the following table:
\begin{table}[ht]
\caption{The summary}
\vspace{0.2cm}
\centering
\begin{tabular}{|c|c|c|}
\hline \hline
Choice of subalgebra & dim$\Mn$ & Properties \\ \hline
$\langle B \rangle $ & $5$ & $R = R_{\mu\nu}^2 = R_{\mu\nu\rho\sigma}^2 = 0 = C_{\mu\nu\rho\sigma}^2$ \\ \hline
$\langle D \rangle $ & $5$ & $R\not = 0$,  $R_{\mu\nu}^2\not = 0$, $ R_{\mu\nu\rho\sigma}^2 \not = 0$ and $C_{\mu\nu\rho\sigma}^2 \not = 0$ \\ 
                                      &                       &               singularity appears as $x_0, x_1, x_B \to \infty$     \\ \hline
$\langle B, D \rangle $ & $4$ & $R = R_{\mu\nu}^2 = R_{\mu\nu\rho\sigma}^2 = 0 = C_{\mu\nu\rho\sigma}^2$ \\ \hline
$\langle B, \alpha_1 D + \alpha_2 K_1 \rangle, $ & $4$ & $R = R_{\mu\nu}^2 = R_{\mu\nu\rho\sigma}^2 = 0 = C_{\mu\nu\rho\sigma}^2$; \\
$\alpha_{1,2} \not = 0 $ &             &                              a (non-trivial) non-reductive coset \\ \hline
$\langle B, D, K_0 \rangle $ & $3$ & Minkowski \\ \hline
\end{tabular}
\label{summary}
\end{table}
\\
Here $R$ denotes the curvature scalar defined by $R = g^{\mu\nu} R_{\mu\nu}$; $R_{\mu\nu}^2 \equiv R^{\mu\nu} R_{\mu\nu}$; 
$R_{\mu\nu\rho\sigma}^2 \equiv R^{\mu\nu\rho\sigma} R_{\mu\nu\rho\sigma}$ and finally the curvature of the Weyl tensor is defined as $C_{\mu\nu\rho\sigma}^2 = C^{\mu\nu\rho\sigma} C_{\mu\nu\rho\sigma}$. The metrics that we obtain 
in this construction (except the Minkowski one) do yield fairly non-trivial Ricci tensor. Thus it is not clear 
to us what matter fields will source such backgrounds. It is therefore not obvious that such matter fields will 
preserve the Galilean Conformal isometry. Thus although our metrics do possess the desired isometry, the full 
background (the metric along with the matter fields sourcing it) may not.

Before we leave this section altogether, a few comments are in order: First, as we remarked earlier in this 
construction we get a family of metrics parametrized by arbitrary real numbers. The redundancy in this description 
does not fix the sign of these parameters and hence does not fix the signature of the metric. However, by assuming the existence of a local orthonormal frame we seem to be able to fix the signature of these metrics and they turn out to be rather non-standard.

Second, note that once we know a metric with the Galilean Conformal isometry in a given dimension, it is 
straightforward to construct a higher dimensional metric with the same isometry by fibering the lower dimensional 
metric over a base manifold
\begin{eqnarray}
ds^2 = f_1(\zeta) d \zeta^2 + f_2 (\zeta) ds_{\rm GCA}^2 \ ,
\end{eqnarray}
where $f_1(\zeta)$ and $f_2(\zeta)$ are two arbitrary functions and $ds_{\rm GCA}^2$ is the metric with the 
Galilean Conformal isometry. This isometry acts non-trivially on the metric $ds_{\rm GCA}^2$ but has no natural 
action on the base manifold. However, a spacetime thus constructed is not a homogeneous space since the Galilean 
Conformal isometry group does not act transitively on the whole manifold. Therefore the homogeneous spaces we 
obtained in $4$ and $5$-dimensions are not related in any obvious manner to the $3$-dimensional Minkowski space 
and are thus truly non-trivial\footnote{We would like to thank J. Simon and J. Figueroa-O'Farill for discussions 
related to this issue.}.

Finally let us return to a point which was made in the initial sections. The $2$d GCA has a degenerate Cartan-Killing form given by
\begin{eqnarray}\label{CKform}
\Omega \sim \begin{pmatrix}
  0& 0 &  -2  & \\
  0 & 1 & 0  & \\
  -2 & 0 & 0  & \\
  &  &  & 0 
\end{pmatrix} \ ,
\end{eqnarray}
where the upper left $3\times 3$ non-degenerate block comes from the $SL(2,R)$ sub-algebra spanned by $\{L^{(\pm)}, L^{(0)}\}$. The rest of the matrix entries are all zeroes.

However, the 2d GCA actually allows for a non-degenerate
2-form over the whole group manifold. The situation is similar to the well-known Nappi-Witten algebra \cite{Nappi:1993ie} (the centrally extended $2$d Euclidean algebra) or the Abelian extension of $d$-dimensional Euclidean algebra considered in {\it e.g.} \cite{Sfetsos:1993rh}. The general construction of an invariant non-degenerate metric for non semi-simple Lie algebra goes by the name of ``double extension" introduced in \cite{FigueroaO'Farrill:1994yf}\footnote{We would like to thank J. Figueroa-O'Farill for explaining this issue to us and bringing this reference to our attention.}. Below we briefly review this.

Let $\hn$ be any Lie algebra and $\hn^*$ be its dual. Let the basis for $\hn$ and $\hn^*$ be respectively denoted by $\{X_a\}$ and $\{X^a\}$ obeying the relation: $\langle X_a, X^b\rangle = \delta_a^b$. Using the fact that $\hn$ acts on $\hn^*$ {\it via} the coadjoint representation, one can define the following Lie algebra structure on the vector space $\hn\oplus \hn^*$
\begin{eqnarray}\label{nonsemi}
\left[X_a, X_b\right] & = & f_{ab}^{\ \ c} X_c \ , \nonumber\\
\left[X^a, X^b\right] & = & 0 \ , \nonumber\\
\left[X_a, X^b\right] & = & - f_{ac}^{\ \ b} X^c \ ,
\end{eqnarray}
where $f_{ab}^{\ \ c}$ are the structure constants for the Lie algebra $\hn$. This defines a semidirect product of $\hn$ and $\hn^*$. It is now possible to define an invariant metric on this semidirect product algebra.

From the definition of the finite $2$d GCA in (\ref{vkmalg2d}) and the Lie algebra structure defined in (\ref{nonsemi}), it is obvious that $X_a \equiv L^{(m)}$ and $X^a \equiv M^{(m)}$, where $m = 0, \pm $. Thus the GCA is isomorphic to the semidirect product of $SL(2,R)$ with its coadjoint representation. We can define a two parameter family of invariant inner products in the following manner:
\begin{eqnarray}
\langle X_a, X_b \rangle = \alpha \Omega_{ab} \ , \quad \langle X_a, X^b \rangle = \beta \delta_a^b \ , \quad \langle X^a, X^b \rangle = 0 \ ,
\end{eqnarray}
where $\alpha$ and $\beta$ are non-zero real numbers and $\Omega_{ab}$ is the non-degenerate Cartan-Killing form for $SL(2,R)$. This construction works for the semidirect 
product of any simple Lie algebra G with its coadjoint representation. It is called the double extension of the 
trivial metric Lie algebra by G \cite{FigueroaO'Farrill:1994yf}.

\section{Summary and conclusions}

In this paper we have systematically constructed metrics in dimensions greater than two which realize the two-dimensional
Galilean Conformal Algebra as their isometry. We classified all the relevant sub-algebras of the 2D GCA and 
in order to construct these metrics looked at a formulation in terms of cosets. Though many choices of these cosets
turned out to produce degenerate metrics, we were able to get some non-trivial higher dimensional metrics. In three
dimensions, we obtained a flat Minkowskian metric which we observed to be the contracted limit of the metric on 
$AdS_3$. In higher dimensions, {\it viz.} four and five, we found several families of metrics, all except one of which
turned out to be Ricci-scalar flat.

It is curious that most of the metrics we have obtained are Ricci-scalar flat. It would be worthwhile trying to understand   
if there is any deeper reason behind this, or if it is a mere co-incidence. One would also like to understand
if there is any fundamental difference between these Ricci-scalar flat metrics and the family which is not, given that they 
were obtained in by similar methods.

Despite the fact that these metrics (except the Minkowski one) seem to have a ``wrong" signature, a further analysis may turn out to be useful in understanding their structure. It will be very interesting to determine the matter fields which source such backgrounds. However since these metrics are neither Lorentzian nor Euclidean, it may be difficult to interpret such ``matter fields" physically.

In the spirit of the gauge/gravity duality, one could look to try and reproduce the correlation functions of the 
2d GCA \cite{Bagchi:2009ca, Bagchi:2009} from a gravity analysis. This might actually be a challenging task as 
there is little chance that modes would separate into normalizable and non-normalizable ones as in the usual AdS case.
But if one is able to perform such computations, then one could claim that these metrics are actually holographically 
dual to the non-relativistic field theories with the GCA as their symmetry algebra.

Another speculation made earlier was that the metrics obtained by this method might realize the infinite dimensional GCA 
as asymptotic symmetries. It has been observed in \cite{Barnich:2006av} that the infinite BMS algebra in three dimensions, which
is isomorphic to the 2d GCA \cite{Bagchi:2010eg}, arises as the asymptotic symmetries of flat space at null infinity.
So, this speculation indeed holds for our construction in three dimensions. The expectation is that the other metrics
which have the finite 2d GCA as their isometries would also realize the infinite GCA in a manner similar to the 
BMS case. In \cite{Compere:2009qm}, following the general scheme of calculating asymptotic symmetries outlined in \cite{Barnich:2001jy},
the authors constructed the asymptotic symmetry algebra for metrics with Schrodinger symmetry and found that 
the infinite extension of the Schrodinger algebra indeed emerges as the asymptotic symmetries of those metrics. 
The obstruction for applying the general formalism of \cite{Barnich:2001jy} to the GCA was the absence of a spacetime
metric. Now that in this work we have derived a number of metrics with the finite GCA as the isometry algebra, it should
in principle be possible to carry out a similar analysis to \cite{Compere:2009qm} and check whether our speculation is indeed correct. 

A natural direction of extending this analysis is to construct the metrics for the higher dimensional GCAs 
by this method of cosets. But the problem of classifying relevant subalgebras quickly becomes intractable and 
the full analysis too unwieldy to attempt by a case-by-case study. This would involve a mathematical 
machinery more elaborate and powerful than what we have used in the two dimensional analysis. 
Another natural extension is to consider the Super-GCA and construction of super-cosets. A natural place to 
begin would be again two dimensions \cite{Mandal:2010gx}. The size of the finite algebra would provide a challenge
which in this case may be overcome by imposing strict ``physical conditions''.
 
To conclude, let us remark on a point we have only fleetingly looked at in this paper. The existence of a 
non-degenerate 2-form on the full finite GCA is an avenue of potential fruitful research. 
Given that there is no field theory known for the GCA, it would be nice to use the construction of Nappi-Witten \cite{Nappi:1993ie} 
and its generalizations \cite{FigueroaO'Farrill:1994yf} to construct a WZW model with the GCA as its symmetry. 
It would also be useful to understand if the infinite extension of the GCA plays any interesting role in this context.

\acknowledgments
It is a pleasure to acknowledge discussions with Paul de Medeiros, Anindya Dey, Jacques Distler, Rajesh Gopakumar, 
Debashis Ghoshal, Joan Simon, Yuji Tachikawa, Masahito Yamazaki and especially Jose Figueroa-O'Farrill. AB would like to thank 
the Institute for Advanced Study, Princeton for their generous hospitality during the initial part of this work. 
AK would like to thank the Institute for Advanced Study for the stimulating environment during the PITP school on
 ``Aspects of Supersymmetry" when a part of this work was carried out. AK is supported by a Simons Postdoctoral 
Fellowship awarded by the Simons Foundation and by the National Science Foundation under Grant Numbers PHY-0969020 
and PHY-0455649.

\appendix

\section{The list of sub-algebras} \label{sub}

Here we enlist the possible choices of the sub-algebras for the finite part of GCA in (\ref{gca2d}). 
We begin by imposing the ``physical" conditions imposed in \cite{SchaferNameki:2009xr} and then relaxing it. 
Just to remind the reader, the ``physical" conditions are:

\noindent 1. $\hn$ does not contain the translation generator $P$.

\noindent 2. $\hn$ contains the boost generator $B$. 

\noindent However we do not impose any constraint on the dimensionality of $\Mn = \Gn / \Hn$.

Let us therefore enlist the possible choices in the descending order in dim$\Mn$:

\noindent (i) dim$\Hn = 1$, dim $\Mn = 5$:
\begin{eqnarray}\label{m5}
\hn = \langle B  \rangle \ , \quad \mn = \langle H, P , D, K_0, K_1 \rangle \ .
\end{eqnarray}
More generally, however we have
\begin{eqnarray}\label{m5gen}
\hn & = & \langle \alpha_1 B + \alpha_2 H + \alpha_3 D + \alpha_4 K_0 + \alpha_5 K_1 \rangle \ ,  \quad \alpha_1 \not = 0 \ , \nonumber\\
\mn & = & \langle H, P, D, K_0, K_1 \rangle \ .
\end{eqnarray}

\noindent (ii) dim$\Hn = 2$, dim$\Mn = 4$:
\begin{eqnarray}\label{m4}
\hn_{(1)} & = & \langle B , K_1 \rangle \ , \quad \mn_{(1)}  = \langle H, P, D, K_0 \rangle  \ , \\
\hn_{(2)} & = & \langle B , D \rangle \ , \quad \mn_{(2)}  = \langle H, P, K_0, K_1 \rangle \ .
\end{eqnarray}
More generally we can have
\begin{eqnarray}\label{m4lc}
\hn_{(3)} & = & \langle B, \alpha_1 D + \alpha_2 K_1 \rangle \ ,  \quad \mn_{(3)} = \langle H, P, D, K_0 \rangle \ , \quad \alpha_2 \not = 0 \ , \\
\hn_{(4)} & = & \langle B, \alpha_1 D + \alpha_2 K_1 \rangle \ ,  \quad \mn_{(4)} = \langle H, P, K_0, K_1 \rangle \ , \quad \alpha_1 \not = 0 \ ,
\end{eqnarray}

\noindent (iii) dim$\Hn= 3$, dim$\Mn = 3$:
\begin{eqnarray}\label{m3}
 \hn_{(1)} & = & \langle B , K_1, \alpha_1 D + \alpha_2 K_0 \rangle \ , \quad \mn_{(1)} = \langle H , P, K_0 \rangle \ , \quad \alpha_1 \not = 0 \ , \\
 \hn_{(2)} & = & \langle B , K_1 ,  K_0 \rangle \ , \quad \mn_{(2)} = \langle H , P, D \rangle  \ . 
\end{eqnarray}
More generally we have
\begin{eqnarray}\label{m3lc}
\hn_{(3)} & = & \langle B, K_1, \alpha_1 D + \alpha_2 K_0 + \alpha_3 K_1 \rangle \ ,  \quad \mn_{(3)} = \langle H, P, K_0 \rangle \ , \quad \alpha_{1, 3} \not = 0 \ , \\
\hn_{(4)} & = & \langle B, K_0, \alpha_1 D + \alpha_2 K_0 + \alpha_3 K_1 \rangle \ ,  \quad \mn_{(4)} = \langle H, P, D \rangle \ , \quad \alpha_{2, 3} \not = 0 \ .
\end{eqnarray}

\noindent (iv) dim$\Hn = 4$, dim$\Mn = 2$:
\begin{eqnarray}
\hn = \langle B, K_1, D, K_0 \rangle \ , \quad \mn = \langle H, P \rangle \ .
\end{eqnarray}

Let us now enlist the possibilities relaxing the condition (2), {\it i.e.} we consider $\hn$ not containing $B$. The choices are:

\noindent (v) dim$\Hn=1$, dim$\Mn=5$:
\begin{eqnarray}
\hn_{(1)} & = & \langle H  \rangle \ , \quad \mn_{(1)} = \langle P, D, K_0, K_1, B  \rangle \ , \\
\hn_{(2)} & = & \langle D  \rangle \ , \quad \mn_{(2)} = \langle H, P, K_0, K_1, B  \rangle \ , \\
\hn_{(3)} & = & \langle K_0  \rangle \ , \quad \mn_{(3)} = \langle H, P, D, K_1, B  \rangle \ , \\
\hn_{(4)} & = & \langle K_1  \rangle \ , \quad \mn_{(4)} = \langle H, P, D, K_0, B  \rangle \ . 
\end{eqnarray}

\noindent \noindent (vi) dim$\Hn = 2$, dim$\Mn=4$:
\begin{eqnarray}
\hn_{(1)} & = & \langle H, D  \rangle \ , \quad \mn_{(1)} = \langle P, K_0, K_1, B  \rangle \ ,  \\
\hn_{(2)} & = & \langle K_0, K_1  \rangle \ , \quad \mn_{(2)} = \langle H, P, D, B  \rangle \ ,  \\
\hn_{(3)} & = & \langle K_0, D  \rangle \ , \quad \mn_{(3)} = \langle H, P, K_1, B  \rangle \ ,  \\
\hn_{(4)} & = & \langle K_1, D  \rangle \ , \quad \mn_{(4)} = \langle H, P, K_0, B  \rangle \ .
\end{eqnarray}
More generally we can have
\begin{eqnarray}
\hn_{(5)}  =  \langle D,  \alpha_1 K_0 + \alpha_2 K_1  \rangle \ ,  \quad \mn_{(5)} & = & \langle H, P, K_1, B \rangle \ .\quad \alpha_1 \not = 0 \ , \nonumber\\
\mn_{(5)} & = & \langle H, P, K_0, B \rangle \ .\quad \alpha_2 \not = 0 \ ,
\end{eqnarray}
\begin{eqnarray}
\hn_{(6)} & = & \langle K_0,  D + \alpha_1 K_1  \rangle \ ,  \quad \mn_{(6)} =  \langle H, P, K_1, B \rangle \ , \\
\hn_{(7)} & = & \langle K_1,  D + \alpha_1 K_0  \rangle \ ,  \quad \mn_{(7)}  =  \langle H, P, K_0, B \rangle \ .
\end{eqnarray}

\noindent (vii) dim$\Hn=3$, dim$\Mn=3$:
\begin{eqnarray}
\hn_{(1)} & = & \langle H, D, K_0  \rangle \ , \quad \mn_{(1)} = \langle P, K_1, B  \rangle \ ,  \\
\hn_{(2)} & = & \langle K_0, K_1, D  \rangle \ , \quad \mn_{(2)} = \langle H, P, B  \rangle \ .
\end{eqnarray}
More generally we can also have
\begin{eqnarray}
\hn_{(3)} = \langle K_0,  D + \alpha_1 K_1, \beta_1 K_0 + \beta_2 K_1  \rangle \ , \quad \mn = \langle H, P, B \rangle  \ .
\end{eqnarray}

For the sake of completeness, below we list the possible sub-algebras relaxing both the conditions (1) and (2):

\noindent (i) dim$\Hn=1$, dim$\Mn=5$:
\begin{eqnarray}
\hn & = & \langle P  \rangle  \ .
\end{eqnarray}

\noindent (ii) dim$\Hn=2$, dim$\Mn=4$:
\begin{eqnarray}
\hn = \langle P, B \rangle \ , \quad \langle P, K_1 \rangle \ , \quad \langle P, H \rangle \ , \quad \langle P, D\rangle \ .
\end{eqnarray}

\noindent (iii) dim$\Hn=3$, dim$\Mn=3$:
\begin{eqnarray}
\hn = \langle P, B, K_1 \rangle \ , \quad \langle P, B, D \rangle \ , \quad \langle P, B, H \rangle \ , \quad \langle P, K_1, D \rangle \ .
\end{eqnarray}
%


\end{document}